\documentclass[a4paper]{article}
\usepackage{graphicx}
\pdfoutput=1
\begin{document}
\title{Calculation of low-energy scattering parameters using
artificial oscillator trap}
\author{
D.V.~Fedorov\footnote{Email: fedorov@phys.au.dk},
and A.M.~Pedersen \\
{\it\small Aarhus University, Aarhus, Denmark}
}
\date{}
\maketitle
	\begin{abstract}
We introduce a recipe to estimate the low-energy scattering
parameters of a quantum few-body system
--- scattering length, effective range, and shape parameter ---
by using only {\em discrete}
state calculations.  We place the system in an artificial oscillator trap
of varying size and calculate the energies of the resulting discrete
states close to the threshold of the system as function of the trap
size. The low-energy scattering parameters are then extracted ---
using a simple analytic formula --- from the functional dependence of
these energies upon the trap size. We first test the recipe against a
simple model problem and then apply it to low-energy nucleon-nucleon
scattering within the nuclear Model with Explicit Mesons in one
sigma-meson approximation.
	\end{abstract}

\section{Introduction}

Explicitly Correlated Gaussians is a popular method for numerical
calculations of quantum few-body systems~\cite{suzuki-varga,bubin,ECGs}.
While offering many advantages, the method also has a drawback -- it
is not {\em directly} applicable to continuum spectrum calculations due
to the specific gaussian asymptotic behaviour of correlated gaussians.
Several approaches have been suggested in the literature to remedy this
shortcoming, including a class of recipes based on the concept of
artificial confining potential, which turns continuum states into
discrete quasi-continuum states. The continuum observables are then
obtained by analysing the calculated quasi-continuum spectrum using one
of many
recipes~\cite{artwell,WCP,r12,rho-osc,zhang,navratil,roper,luu,busch,bazak}.

Here we introduce yet another artificial confining potential method to
estimate the low-energy scattering parameters of a few-body system using
only discrete spectrum calculations.  We place the system in a suitable
oscillator trap and calculate the energies of discrete states closest to
the threshold of the system as function of the trap size. The low-energy
scattering parameters are then extracted from the functional dependence
of these energies upon the trap size.

What sets our recipe apart from the existing ones is that~i) we explicitly
target the low-energy parameters (scattering length, effective range, and
shape-parameter) and~ii) we extract the low-energy parameters directly
from the calculated energies by fitting a simple analytic formula to
their functional dependence upon the trap size.

The recipe is based on the assumption that if two interaction models
provide similar energy levels in an artificial trap, their scattering
parameters must also be similar.  In our recipe one of the interaction
models is the zero-range potential model, implemented as a boundary
condition at the origin, for which an analytic formula for the spectrum
in the oscillator trap can be derived.  The low-energy parameters of any
other interaction model can then be extracted by fitting its calculated
spectrum to the zero-range formula.

We first test the recipe against a simple model --- the Volkov
potential --- and then apply it to calculate the triplet nucleon-nucleon
scattering length and effective range within the nuclear Model with
Explicit Mesons in one sigma-meson approximation.

\section{Harmonic oscillator with zero-range potential}

The problem of a quantum harmonic oscillator with a zero-range
potential at the origin has been investigated in the past in several
publications~\cite{luu,busch,suzuki,yip}. Nevertheless, we reproduce here
the pertinent formulae for completeness and because previous publications
fell short of the shape parameter that we need.

The $s$-wave Schrödinger equation for a particle with mass $m$ in an
oscillator trap with frequency $\omega$ and a short-range potential
$V(r)$ is given as
	\begin{equation}\label{eq-ozr}
\left(
-\frac{\hbar^2}{2m}\frac{d^2}{dr^2}+V(r)+\frac12 m\omega^2 r^2-E
\right)u(r)=0 \;,
	\end{equation}
where $r$ is the radial coordinate, $u(r)$ is the radial\footnote{The
radial wavefunction $u(r)$ is given as $u(r)=\frac{\psi(r)}{r}$ where
$\psi(r)$ is the ordinary wavefunction.} wavefunction and $E$ is the
energy of the particle.

If $V(r)$ is a {\em zero}-range potential --- with scattering length~$a$,
effective range~$r_e$ and shape parameter~$P$ --- then it completely
disappears from
the equation and instead changes the boundary condition at the
origin: from the usual condition,
	\begin{equation}
\left.{u}\right|_{r=0}=0 \,,
	\end{equation}
to the zero-range boundary condition~(see Appendix~\ref{sec-zrp}),
	\begin{equation}\label{eq-zrp}
\left. \frac{u'}{u} \right|_{r=0} = \frac1{a} +\frac12 r_e
\frac{2mE}{\hbar^2} +P r_e^3 \left(\frac{2mE}{\hbar^2}\right)^2 \,.
	\end{equation}

The solution to equation~(\ref{eq-ozr}), regular at infinity,
is given by the parabolic cylinder function~$U(\alpha,z)$,
	\begin{equation}
u(r) \propto U\left(-\frac{E}{\hbar\omega},\frac{r}{b_o}\right) \,,
	\end{equation}
where $b_o=\sqrt{\hbar/(2m\omega)}$ is the trap range.  Inserting
this into the boundary condition Eq.~(\ref{eq-zrp}) gives
	\begin{equation}
\frac1{b_o}\frac{U'\left(-\frac{E}{\hbar\omega},0\right)}
{U\left(-\frac{E}{\hbar\omega},0\right)}
=\frac1{a}+\frac12\frac{r_e}{b_o^2}\frac{E}{\hbar\omega}
+P\frac{r_e^3}{b_o^4}\left(\frac{E}{\hbar\omega}\right)^2 \,.
	\end{equation}
It can be shown (see Appendix~\ref{sec-parabolic}) that
	\begin{equation}
\frac{U'(\alpha,0)}{U(\alpha,0)}=
-\sqrt{2}
\frac{\Gamma(\frac{\alpha}{2}+\frac34)}
{\Gamma(\frac{\alpha}{2}+\frac14)} \,,
	\end{equation}
which gives the following analytic eigenvalue equation for a particle
in a combined zero-range and oscillator potential,

	\begin{equation}\label{EQN}
-\sqrt{2}
\frac{\Gamma(-\frac{E}{2\hbar\omega}+\frac34)}
{\Gamma(-\frac{E}{2\hbar\omega}+\frac14)}
=\frac{b_o}{a}+\frac12\frac{r_e}{b_o}\frac{E}{\hbar\omega}
+P\left(\frac{r_e}{b_o}\right)^3\left(\frac{E}{\hbar\omega}\right)^2 \,.
	\end{equation}

For the given $a$, $r_e$, $P$, $\omega$ the solutions
$E(a,r_e,P,\omega)$ to this equation are the eigen-energies of
the corresponding zero-range potential in the oscillator
trap with frequency~$\omega$ and range
$b_o=\sqrt{\hbar/(2m\omega)}$. The limit $a{\to}0$ recovers the free
$s$-wave oscillator spectrum $(2n+3/2)\hbar\omega$.

Alternatively, given an eigen-energy $E(b_o)$ of any interaction model as
function of the oscillator range $b_o$ (at least three distinct values),
this equation provides the corresponding low-energy scattering parameters
$a$, $r_e$, and $P$.  And that is the essence of our recipe.

\section{Test against simple model}
As a test of the introduced recipe we consider a particle in
the Volkov potential~\cite{volkov},
	\begin{equation}
W(r)=V_R e^{-r^2/b_R^2} + V_A e^{-r^2/b_A^2} \,,
	\end{equation}
where $V_R=144.86$~MeV, $b_R=0.82$~fm,
$V_A=-83.34$~MeV, $b_A=1.60$~fm. The mass of the particle --- approximately
the reduced mass of two nucleons ---
is given as~\cite{volkov},
	\begin{equation}
m=\frac12\frac{(\hbar c)^2}{41.47 \mathrm{~MeV~fm^2}}
=469.471 \mathrm{~MeV} \,.
	\end{equation}
The reported scattering length\footnote{Here we use the opposite sign
convention for the scattering length as compared to~\cite{volkov}.}
for this system is $a=-10.082$~fm and the bound state
energy~$E_b=-0.54592$~MeV~\cite{volkov}.

\begin{figure}
\caption{Solid dots: the two lowest energy levels of a particle
in the Volkov potential and an oscillator trap with range~$b_o$,
Eq.~(\ref{eq-wo}), as function of the trap range.  Solid lines: the
corresponding energies of the same particle in the same trap but with
the zero-range potential (ZRP), Eq.~(\ref{EQN}), with $a=10.08$fm,
$r_e=2.38$fm, and $P=0.028$.  The dashed line indicates the bound
state energy as calculated in~\cite{volkov}.  }

\centerline{\includegraphics{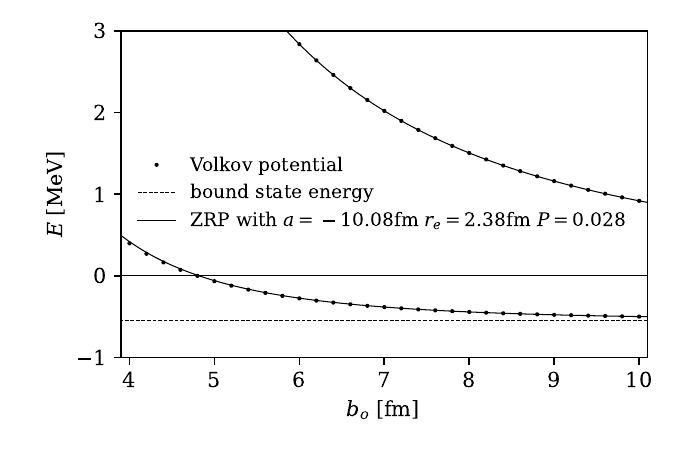}}
\label{fig-volkov}
\end{figure}

According to our prescription we now add an artificial oscillator trap
with the range $b_o$ to the system. The corresponding Schrödinger
equation is then given as
	\begin{equation}\label{eq-wo}
\left[-\frac{\hbar^2}{2m}\frac{\partial^2}{\partial r^2}+W(r)
+\frac14\left(\frac{\hbar^2}{2mb_o^2}\right)\frac{r^2}{b_o^2}
\right]u(r)=Eu(r) \,.
	\end{equation}
The numerically calculated spectrum of this system for different values of
$b_o$ is shown on Fig.~(\ref{fig-volkov}) as black dots. The energies of
the ground state and the first excited state, collected from the last two
femtometers, were used to estimate the scattering parameters by fitting
to equation Eq.~(\ref{EQN}). The result is $a=-10.08$fm, $r_e=2.38$fm,
and $P=0.028$.

\begin{figure}
\caption{Convergence with with respect to trap range: the low-energy
parameters of the Volkov potential estimated in a trap with range
$b_o$. The dots are the calculated parameters, the lines are only for
guiding the eye.}

\centerline{\includegraphics{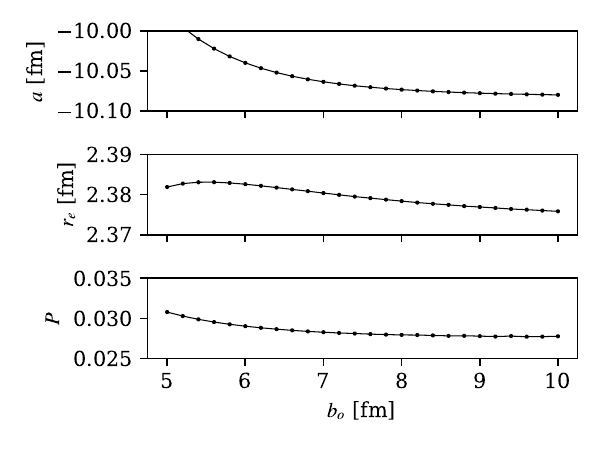}}
\label{fig-volkov-conv}
\end{figure}

To illustrate the agreement between the two interaction models (Volkov
and zero-range) the figure also shows --- with solid lines --- the
eigen-energies of the zero-range potential with those low-energy
parameters.  The agreement seems very good and the results seem to have
converged within four decimal digits for the scattering length
(Fig.~(\ref{fig-volkov-conv}) illustrates the convergence of the results
with respect to the trap size).

\section{Neutron-proton scattering in $\sigma$MEM}

In the nuclear model with explicit mesons (MEM) the nucleons interact by
(literally) emitting and absorbing mesons~\cite{MEM}. In its simplest
version, $\sigma$MEM, a pair of nucleons can emit and absorb a (scalar
isoscalar) sigma-meson. It has been shown that this version of MEM
is able to provide a bound neutron-proton state, the deuteron, with a
reasonable binding energy and radius. However the question remains
about the corresponding scattering observables -- the scattering length
and the effective range. In the following we calculate those observables
using the introduced prescription.

In $\sigma$MEM the physical neutron-proton system is a
superposition of two subsystems: the neutron-proton subsystem, and the
neutron-proton-sigma subsystem. The total wavefunction, $\Psi_{np}$,
is a two-component structure,
       \begin{equation}\label{eq-psinp}
\Psi_{np}=\left(
\begin{array}{c}
\psi_{np} \\
\psi_{np\sigma}
\end{array}
\right) \,,
	\end{equation}
where $\psi_{np}$ is the wavefunction of the neutron-proton subsystem and
$\psi_{np\sigma}$ is the wavefunction of the neutron-proton-sigma
subsystem.
The corresponding Hamiltonian is a block-matrix operator,
       \begin{equation}\label{eq-1}
H=\left(
\begin{array}{cc}
K_n+K_p+V_\mathrm{trap} & W^\dagger  \\
W & K_n+K_p+K_\sigma+m_\sigma \\
\end{array}
\right)
\;,
       \end{equation}
where $K_n$, $K_p$, $K_\sigma$ are kinetic energy operators for the
neutron, proton, and the sigma-meson, $m_\sigma$ is the mass of the
$\sigma$-meson, $V_\mathrm{trap}$ is the artificial confining potential,
and $W$/$W^\dagger$ is the sigma-meson
generation/annihilation operator taken from~\cite{MEM},
       \begin{equation}
W=20.35\mathrm{MeV}\,\times\,
\exp\left(-\frac{r_{np}^2+r_{np\sigma}^2}{(3\mathrm{fm})^2}\right) \,,
       \end{equation}
where $r_{np}$ is the distance between the neutron and the proton,
and $r_{np\sigma}$ is the distance between the sigma-meson and the
neutron-proton centre of mass.

The $np\sigma$ subsystem does not need a trap since at the energies
involved---just above the $np$-threshold---this subsystem is in the
classically forbidden region and is therefore confined by itself.

Each component of the total wavefunction is expanded in correlated
gaussians, the parameters of which are optimised by minimising the sum of
the ground and the first excited state of the system.  Experimentations
show that as few as 4~gaussians in the $np$-channel and 12~gaussians in
the $np\sigma$-channel are sufficient to provide the results that are
converged to within three decimal digits.

\begin{figure}
\caption{Solid dots: the two lowest energy levels of the neutron-proton
system in an oscillator trap with range $b_o$ calculated in
$\sigma$MEM. Solid lines: the corresponding energies of the neutron-proton
system in the same oscillator trap interacting via a zero-range
potential (ZRP) with the given parameters . The dashed line indicates
the binding energy as calculated in~\cite{MEM}. }

\centerline{\includegraphics{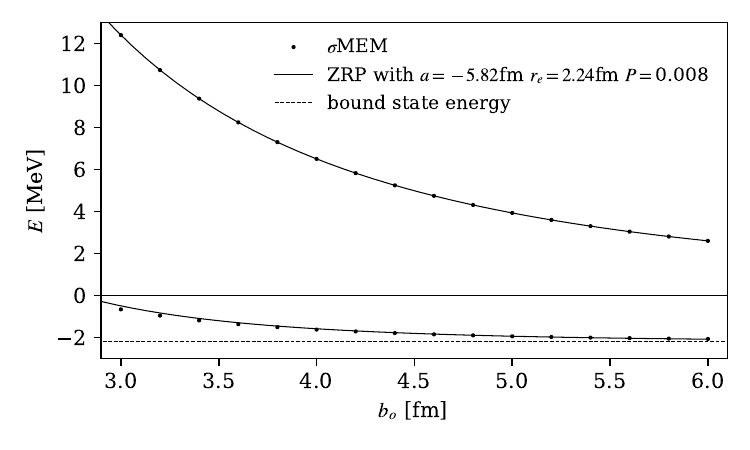}}
\label{fig-D}
\end{figure}

\begin{figure}
\caption{Convergence with respect to the trap range: the $\sigma$MEM
low-energy neutron-proton parameters estimated in a trap with
range~$b_o$. The dots are the calculated parameters, the lines are only
for guiding the eye.}

\centerline{\includegraphics{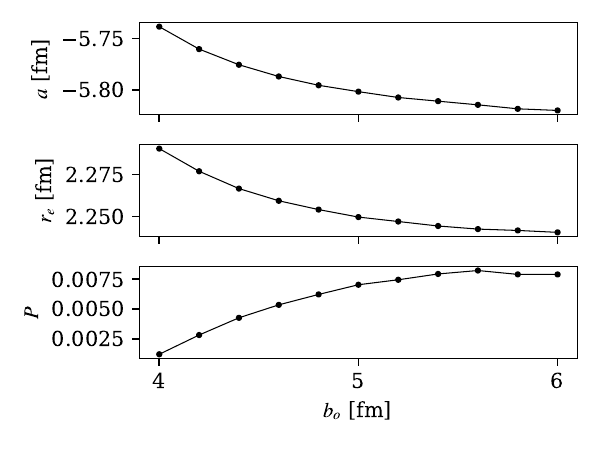}}
\label{fig-D-conv}
\end{figure}

The two lowest calculated eigen-energies of the Hamiltonian
Eq.~(\ref{eq-1}) are shown on Fig.~(\ref{fig-D}) together with the
energies of a zero-range potential with parameters $a=5.82$fm,
$r_e=2.24$fm, and $P=0.008$.  The convergence of the calculated
scattering parameters with respect to trap size is illustrated on
Fig.~(\ref{fig-D-conv}).

As in the test case, the zero-range model fits the data very
well. However the resulting scattering length and
effective range --- while qualitatively correct ---
overestimate the experimental values ($a=5.41$fm,
$r_e=1.74$fm~\cite{a-re}) indicating that the $\sigma$MEM model is not
complete.  That is expected however as any meson exchange interaction
model should probably include the pion exchange contribution (which
$\sigma$MEM lacks).

\section{Conclusion}
We have introduced a new recipe to estimate the low-energy scattering
parameters --- scattering length, effective range, and shape-parameter
--- of a few-body quantum system using only {\em discrete} spectrum
calculations.

The recipe belongs to the class of artificial confining potential recipes.
It involves placing the few-body system in a suitable artificial harmonic
oscillator trap which turns continuum scattering states into discrete
states. The energies of several discrete states close to the threshold of
the system are calculated as function of the trap size.  The low-energy
scattering parameters are then extracted from the functional dependence
of the energies upon the trap size.

Unlike the existing recipes from the same class we i)~explicitly target
the low-energy parameters and ii)~extract the parameters directly from
the calculated energies by fitting an analytic formula to their functional
dependence upon the trap size.

We have first tested the recipe against a simple model and then applied it
to calculate the low-energy neutron-proton scattering parameters within
the nuclear Model with Explicit Mesons in one sigma-meson approximation.
The results show that this approximation --- although qualitatively
correct --- overestimates the scattering parameters by about 10\% to
20\%, indicating that the approximation is incomplete, most probably
due to its lack of the pion exchange contribution.  The latter is the
subject of the ongoing investigation.

\appendix

\section{Zero-range potential as boundary condition}\label{sec-zrp}
Consider the $s$-wave radial Schrödinger equation for
a particle in a short-range potential $V(r)$,
	\begin{equation}\label{eq-short}
\left(
-\frac{\hbar^2}{2m}\frac{\partial^2}{\partial r^2}+V(r)-E
\right)u(r)=0 \;,
	\end{equation}
where $r$ is the radial coordinate, $m$ is the mass of the particle,
$u(r)$ is the radial\footnote{The radial wavefunction $u(r)$ is given as
$u(r)=\frac{\psi(r)}{r}$ where $\psi(r)$ is the ordinary wavefunction.}
wavefunction and $E$ is the energy of the particle.
The short-range potential $V(r)$ is assumed to have a certain (short)
range $b$ beyond which the potential is negligible.  This range divides
the space into two regions: the inside region, $r<b$, where the potential
is significant, and the outside region, $r>b$, where the potential is
negligible.  The logarithmic derivatives of the solutions in the inside
and outside regions must be matched at $r=b$.

In the outside region, $r>b$, where the potential is negligible, the
solution to the equation Eq.~(\ref{eq-short}) is a free wave given as
	\begin{equation}
u_\mathrm{out}(r) = \sin(kr+\delta) \,,
	\end{equation}
where $h^2k^2/(2m)=E$ and $\delta$ is the phase-shift.
The logarithmic derivative of this solution at $r=b$,
	\begin{equation}\label{eq-log}
\left. \frac{u_\mathrm{out}'}{u_\mathrm{out}} \right|_{r=b}
= k\cot(kb+\delta) \,,
	\end{equation}
must be matched with the solution from
the inside region.

Now, the {\em zero-range} potential amounts to moving the condition
Eq.~(\ref{eq-log}) to the origin,
	\begin{equation}
\left. \frac{u_\mathrm{out}'}{u_\mathrm{out}} \right|_{r=0}
= k\cot(\delta) \,.
	\end{equation}
This approximation might be expected to be valid in the low-energy limit
where $kb\to 0$.  In the same limit one customarily defines
	\begin{equation}
k\cot(\delta)=\frac1{a}+\frac12 r_e k^2+Pr_e^3 k^4+O(k^6) \,,
	\end{equation}
where $a$ is the scattering length, $r_e$ is the effective range, and $P$
is the shape parameter. Combining the last two equations reduces
the zero-range potential to the following energy-dependent boundary
condition at the origin,
	\begin{equation}
\left. \frac{u'}{u} \right|_{r=0} = \frac1{a}
+\frac12 r_e \frac{2mE}{\hbar^2}
+P r_e^3 \left(\frac{2mE}{\hbar^2}\right)^2 \,.
	\end{equation}

The zero-range potential can be thought of as a specific limit of a
short-range potential where the potential depth tends to infinity while
the range approaches zero such that the effective range parameters remain
unchanged. Consequently any additional potential that is finite at the
origin (in particular the oscillator potential, which is zero at the
origin) will not modify the boundary condition for the (infinitely deep
and vanishingly short) zero-range potential.

\section{Parabolic cylinder function}\label{sec-parabolic}
The parabolic cylinder function $U(a,z)$ is the solution to the
differential equation
	\begin{equation}
\left(-\frac{d^2}{dz^2}+\frac14 z^2 + a\right) u(z) = 0
	\end{equation}
with the asymptotics
	\begin{equation}
U(a,z)\sim z^{-a-\frac12}e^{-\frac14 z^2} \,.
	\end{equation}
Its integral representations are given as~\cite{functions},
	\begin{equation}
U(a,z)=\frac{e^{-\frac14 z^2}}{\Gamma\left(a+\frac12\right)}
\int_0^\infty e^{-zt}t^{a-\frac12}e^{-\frac12 t^2}dt
\,,\; \Re a>-\frac12 \;,
	\end{equation}
	\begin{equation}
U(a,z)=\sqrt{\frac2{\pi}}e^{\frac14 z^2}
\int_0^\infty \cos\left(zt+\frac{\pi}{2}a+\frac{\pi}{4}\right)
t^{-a-\frac12}e^{-\frac12 t^2}dt
\,,\; \Re a<\frac12 \;.
	\end{equation}
From these representations one can directly calculate the following
values at~$z{=}0$,
	\begin{equation}
U(a,0)=\frac{\sqrt{\pi}\,2^{-\frac{a}{2}-\frac14}}
{\Gamma\left(\frac{a}{2}+\frac34\right)} \,,\;
U'(a,0)=-\frac{\sqrt{\pi}\,2^{-\frac{a}{2}+\frac14}}
{\Gamma\left(\frac{a}{2}+\frac14\right)} \,.
	\end{equation}

\end{document}